\documentclass[conference]{IEEEtran}
\IEEEoverridecommandlockouts

\usepackage{cite}
\usepackage{amsmath,amssymb,amsfonts}
\usepackage[ruled,norelsize,linesnumbered]{algorithm2e}
\usepackage{graphicx}
\usepackage{textcomp}
\usepackage{xcolor}
\usepackage{graphicx}
\usepackage[caption=false,font=footnotesize]{subfig}
\def\BibTeX{{\rm B\kern-.05em{\sc i\kern-.025em b}\kern-.08em
		T\kern-.1667em\lower.7ex\hbox{E}\kern-.125emX}}

\newcommand{\mcal}[1]{\mathcal{#1}}
\newcommand{\mb}[1]{\mathbf{#1}}
\newcommand{\real}[1]{\mathbb{R}^{#1}}
\newcommand{\prob}[1]{\mathbb{P} \left[ #1 \right]}
\newtheorem{theorem}{Theorem}
\newcommand{\lap}[1]{\text{Lap}\left( #1 \right)}
\newcommand{\ldef}{:=}
\newcommand{\spnorm}[2]{\left\lVert #1 \right\rVert_{#2}}
\newcommand{\ess}{\text{ESS}}
\newcommand{\load}{\text{L}}

\DeclareMathOperator*{\argmax}{argmax}
\DeclareMathOperator*{\argmin}{argmin}
\newcommand{\lr}{\mathcal{LR}}
\newcommand{\stack}[2]{\text{\texttt{stack}}\left( #1 , #2 \right)}
\newcommand{\tth}[1]{{#1}^{\text{th}}}
\makeatletter
\newcommand{\removelatexerror}{\let\@latex@error\@gobble}
\makeatother
\makeatletter
\newcommand{\nosemic}{\renewcommand{\@endalgocfline}{\relax}}% Drop semi-colon ;
\newcommand{\dosemic}{\renewcommand{\@endalgocfline}{\algocf@endline}}% Reinstate semi-colon ;
\let\oldnl\nl% Store \nl in \oldnl
\newcommand{\nonl}{\renewcommand{\nl}{\let\nl\oldnl}}% Remove line number for one line
\makeatother
\begin{document}
	\title{Differentially Private Load Restoration for Microgrids with Distributed Energy Storage \\
		\thanks{This work was supported in part by the Faculty Research Grant and the Chancellor's Fellowship of UC Santa Cruz, Seed Fund Award from CITRIS and the Banatao Institute at the University of California, and the Hellman Fellowship.}
	}
	\author{\IEEEauthorblockN{Shourya Bose and Yu Zhang}
		\IEEEauthorblockA{\textit{Department of Electrical and Computer Engineering} \\
			\textit{University of California, Santa Cruz}\\
			\texttt{\{shbose,\,zhangy\}@ucsc.edu}}
	}
	\maketitle
	\begin{abstract}
		Distributed energy storage systems (ESSs) can be efficiently leveraged for load restoration (LR) for a microgrid (MG) in island mode. When the ESSs are owned by third parties rather than the MG operator (MGO), the ESS operating setpoints may be considered as private information of their respective owners. Therefore, efforts must be put forth to avoid the disclosure through adversarial analysis of load setpoints. In this paper, we consider a scenario where LR takes place in a MG by determining load and ESS power injections through the solution of an AC optimal power flow (AC-OPF) problem. Since the charge/discharge mode at any given time is assumed to be private, we develop a differentially-private mechanism which restores load while maintaining privacy of ESS mode data. The performance of the proposed mechanism is demonstrated for a $33$-bus MG.
	\end{abstract}
	\section{Introduction}
	Power distribution systems face constant risk of outage due to various factors~\cite{HA-etal:2019} such as extreme weather conditions, system misconfiguration, overloading, etc. In this context, effective load restoration (LR) for electricity end users becomes an important but challenging task. A specific type of power network suitable for this purpose is the microgrid (MG). It is a geographically localized distribution system containing both loads and distributed energy resources  (DERs). Federated control of MGs enables automatic LR when external supply of power is interrupted. A key element of a MG is the microgrid controller (MGC), which is a central computer that controls loads and DERs. Upon loss of external power, the MGC needs to provide new setpoints for all controllable and networked elements within the MG. If the setpoints are chosen appropriately, desired LR performance can be achieved~\cite{GR-etal:2018}.
	
	MG operations are fundamentally distributed processes that are coordinated by the MGC. It is of paramount importance to ensure privacy of the data for all participating users. Some examples of user data include voltage and current values, operating modes and setpoints, a list of appliances being operated, etc. Leakage of private data during MG operation can lead to loss of trust, user attrition, and liabilities on part of a microgrid operator (MGO), thereby posing a threat to increased adoption of MGs for ensuring security and reliability of future power grids. The book~\cite{KGB-MHA-SSI:2017} is an in-depth reference for various privacy concerns in MGs and smart grids.
	
	Conventional distributed approaches for MG operation such as primal-dual gradient methods~\cite{QZ-YG-ZW-FB:2021} and model-free machine-learning methods~\cite{QZ-KD-ZW-QH:2020} provide a certain degree of privacy thanks to the locality of data. However, those methods do not guarantee that adversarial agents with strong computational capabilities cannot infer private data from local information such as voltage values. To overcome this shortcoming, we utilize the framework of \emph{differential privacy} (DP)~\cite{CD-etal:2006} to address data privacy in MG operation. DP posits that in order to protect the identity of individuals in a dataset, a carefully calibrated noise should be added to the response of any query on the dataset. The noise parameters are chosen to ensure that the noisy response makes it difficult to infer the presence or absence of an individual's data in the dataset. Since DP is an \emph{information theoretic} concept, it is not vulnerable to adversaries with massive computational power.
	\subsubsection*{Literature Review} 
	Extensive efforts have been carried out on the topic of LR in MGs. Existing approaches include feedforward control of distributed generation~\cite{JYP-etal:2021}, risk-limiting approach~\cite{FS-etal:2020}, robust model-predictive control~\cite{SC-etal:2020}, bi-level optimization approach~\cite{QZ-ZM-YZ-ZW:2021}, as well as reinforcement learning based control~\cite{SY-etal:2020}. Comparatively, DP has received little attention as a framework for analyzing privacy in power systems. \cite{SH-UT-GJP:2017} considers differentially private distributed optimization to schedule charging of electrical vehicles. \cite{FZ-JA-SHL:2019} investigates differential privacy of load data against aggregation queries. \cite{VD-etal:2021} develops a differentially private DC-OPF solver which exploits the structure of the underlying optimization problem to ensure privacy of load setpoints. However, privacy of generation data in a LR setting, which is indispensable for MGs with multiple producers or prosumers, has not received enough study. Especially, privacy issues have not analyzed for distribution systems modeled with the nonlinear AC Power Flow (AC-PF) equations.
	\subsubsection*{Contributions}
	In this work, we formulate a load restoration problem for a MG equipped with energy storage systems (ESSs) over multiple time steps. The charge/discharge mode of each ESS on every time step is treated as a private datum, for which we leverage DP to obfuscate its value. The resulting load setpoints may not be feasible, and we propose a post-processing procedure to restore feasibility while ensure monotonicity in load restoration. Finally, performance of the proposed approach is demonstrated through extensive simulations tested on a $33$-bus MG.

	\subsubsection*{Notation}
	The symbols $\real{}$ and $\real{}_{+}$ denote the set of real numbers and non-negative real numbers, respectively. The notation $[N]$ denotes the set $\{1,\cdots,N\}$. The probability of event $A$ is given by $\prob{A}$. Vectors and matrices are denoted in boldface. $\mb{0}_m$ and $\mb{1}_m$ represent the $m$-dimensional vector of all zeros and ones. $(\mb{a})_j$ is the $\tth{j}$ element of vector $\mb{a}$. The condition $\mb{a}\sim_1 \mb{b}$ is true if and only if these two vectors differ only at one entry (i.e. $\spnorm{\mb{a}-\mb{b}}{0}=1$). For a sequence of vectors  $\{\mb{a}_k\}_{k\in\mcal{K}}$, the expression $\stack{\mb{a}_k}{\mcal{K}}$ represents the stacked vector $[\mb{a}_1^\top,\mb{a}_2^\top,\cdots,\mb{a}_{|\mcal{K}|}^\top]^\top$.
	
	\nopagebreak
	\section{Problem Statement}
	\nopagebreak
	\subsection{Differential Privacy}
	In this section, we briefly describe DP that ensures privacy of datasets manipulated by an algorithm. Let $\mcal{D}$ denote the data space of a single user. A \emph{dataset} $\mb{d}\in\mcal{D}^n$ contains data of $n$ users. A \emph{query} $\mcal{F}:\mcal{D}^n \mapsto \real{k}$ provides $k$ `responses' based on a given dataset. A \emph{mechanism} $\mcal{M}$ is a stochastic function of a query, i.e. $\mcal{M}(\mb{d}) = g(\mcal{F}(\mb{d}),\mb{w})$, where $\mb{w}$ is a random vector drawn from a specific distribution. A common class of mechanisms are ones which simply add noise to the query output, i.e. $\mcal{M}(\mb{d}) \ldef \mcal{F}(\mb{d})+\mb{w}$, to which we restrict our attention. A mechanism $\mcal{M}$ is said to be $\varepsilon$-differentially private ($\varepsilon$-DP) if for some $\varepsilon>0$,
	\begin{gather*}
		\prob{\mcal{M} (\mb{d}) \in U} \leq e^\varepsilon \prob{ \mcal{M}(\mb{d}') \in U}\\
		\forall U \subseteq \real{k}, \;\; \forall \mb{d},\mb{d}'\in\mcal{D}^n \text{ such that } (\mb{d}\sim_1\mb{d}').
	\end{gather*}
	Note that smaller values of $\varepsilon$ ensure higher privacy (see details in Section 2.3 of~\cite{CD-AR:2014}). 
	% To see this, note that it follows from the symmetry of $\mb{d}$ and $\mb{d}'$ in the definition of $\varepsilon$-DP that
	%\begin{align*}
	%	e^{-\varepsilon} \prob{\mcal{M}(\mb{d}') \in U} \leq \prob{\mcal{M} (\mb{d}) \in U} \leq e^\varepsilon \prob{ \mcal{M}(\mb{d}') \in U}
	%\end{align*}
	%and as $\varepsilon \rightarrow 0$, the events $\mcal{M}(\mb{d})\in U$ and $\mcal{M}(\mb{d}')\in U$ occur with almost similar probabilities for every $U$ and are therefore indistinguishable. 
	As given by the following theorem, post-processing preserves DP.
	\begin{theorem}[Prop. 2.1,~\cite{CD-AR:2014}]
		\label{th:pp}
		Let $\mcal{M}:\mcal{D}^n \mapsto \real{k}$ be an $\varepsilon$-DP mechanism, and let $\mcal{G}:\real{k}\mapsto \real{m}$ be an arbitrary randomized mapping. Then, $\mcal{G} \circ \mcal{M}$ is $\varepsilon$-DP.
	\end{theorem}
	
	Based on the Laplace distribution, the popular Laplace mechanism is $\varepsilon$-DP (cf. Thm. 3.6,~\cite{CD-AR:2014}). A centered Laplace random variable $X\sim \lap{b}$ ($b>0$) has the probability density function $f_X(x) = \frac{1}{2b}e^{-\frac{|x|}{b}}, \; \forall x \in \real{}$. $X$ is zero mean and has variance $2b^2$. The Laplace mechanism $\mcal{M}(\mb{d}) = \mcal{F}(\mb{d}) + \mb{w}$ selects each noise element $(\mb{w})_i$ independently and identically distributed (i.i.d.) such that $(\mb{w})_i \sim \lap{\frac{\Delta}{\varepsilon}}$, where $\Delta$ is the $l_1$-sensitivity of the query, defined as
	\begin{align*}
		\Delta \ldef \max\limits_{\mb{d},\mb{d}'\in\mcal{D}^n: \, (\mb{d}\sim_1 \mb{d}')} \spnorm{\mcal{F}(\mb{d})-\mcal{F}(\mb{d}')}{1}.
	\end{align*}
	\begin{theorem}[Thm. 3.6,~\cite{CD-AR:2014}]
		\label{th:lap-dp}
		The Laplace mechanism is $\varepsilon$-DP.
	\end{theorem}
	% 	\nopagebreak
	
	\subsection{Load Restoration Problem}
	We consider an island-mode MG with multiple ESSs and loads. The MG can be represented as a directed graph $\mcal{G} = (\mcal{N},\mcal{E})$ where $\mcal{N}\ldef[N]$ denotes the set of \emph{nodes} or \emph{buses}, and $\mcal{E}\ldef [E]$ denotes the set of \emph{branches} or \emph{lines}. We assume that the MG has a \emph{radial} topology, i.e. $\mcal{G}$ is a tree. $\mcal{N} = \mcal{N}^\ess \cup \mcal{N}^\load$, wherein the two disjoint sets $\mcal{N}^\ess$ and $\mcal{N}^\load$ represent ESS and load buses, respectively. We denote the sizes of these sets as $N_E\ldef |\mcal{N}^\ess|$ and $N_L \ldef |\mcal{N}^\load|$ (with $N=N_L+N_E$). For the scenario of a load attached to an ESS, it can be modeled by allocating them adjacent buses connected by an impedance-free line. To focus on privacy of ESS data, we do not include distributed generation sources in the MG, although they can be easily incorporated into our model. 
	
	The LR problem is formulated over a time horizon $\mcal{T}\ldef [T]$, and the variable $t\in\mcal{T}$ indicates the time step. On time step $t$, $\mb{p}_t \in \real{N}$ and $\mb{q}_t \in \real{N}$ denote real and reactive bus injections while $\mb{P}_t\in\real{E}$ and $\mb{Q}_t\in\real{E}$ are real and reactive line flows. $\mb{v}_t\in\real{N}_{+}$ and $\mb{l}_t\in\real{E}_{+}$ denote squared magnitude of nodal voltage phasors and line current phasors, respectively.  $\mb{s}_t\in\real{N_E}$ denotes the state-of-charge (SoC), and $\mb{d}_t\in\{0,1\}^{N_E}$ denotes the discharge/charge mode of each ESS, where
	$0$ indicates discharge while $1$ indicates charge mode. We maintain the convention that power injected into the MG has a positive sign, while power withdrawn from the MG has a negative sign.

	We collect the aforementioned variables over the entire time horizon as $\mb{p}\ldef\stack{\mb{p}_t}{\mcal{T}}$, $\mb{q}\ldef\stack{\mb{q}_t}{\mcal{T}}$, $\mb{P}\ldef\stack{\mb{P}_t}{\mcal{T}}$, $\mb{Q}\ldef\stack{\mb{Q}_t}{\mcal{T}}$, $\mb{v}=\stack{\mb{v}_t}{\mcal{T}}$, $\mb{l}\ldef\stack{\mb{l}_t}{\mcal{T}}$, $\mb{s} \ldef \stack{\mb{s}_t}{[T+1]}$, and the \emph{mode vector} $\mb{d} \ldef \stack{\mb{d}_t}{\mcal{T}}$.
	The voltages, currents, injections and branch flows on every time step are  related through a convex second-order cone (SOC) relaxation of the nonconvex \emph{DistFlow} AC-PF equations. Under mild conditions, such a relaxation is exact for radial networks (cf. Thm. 1,~\cite{LG-etal:2015}). For any time step $t\in\mcal{T}$, the equations are given as
	\begin{subequations}
		\label{eq:DF}
		\begin{gather}
			\label{eq:DF1}
			\sum_{\substack{i\in\mcal{N}:\\b = (i,j)\in\mcal{E}}} (\mb{P}_t)_b = (\mb{p}_t)_j + \sum_{\substack{k\in\mcal{N}:\\b'=(j,k)\in\mcal{E}}} (\mb{P}_i)_{b'} - r_{b'}(\mb{l}_t)_{b'} \\
			\label{eq:DF2}
			\sum_{\substack{i\in\mcal{N}:\\b = (i,j)\in\mcal{E}}} (\mb{Q}_t)_b = (\mb{q}_t)_j + \sum_{\substack{k\in\mcal{N}:\\b'=(j,k)\in\mcal{E}}} (\mb{Q}_i)_{b'} - x_{b'}(\mb{l}_t)_{b'} \\
			%\notag
			\label{eq:DF3}
			(\mb{v}_t)_i - (\mb{v}_t)_j = 2(r_b(\mb{P}_t)_b + x_b (\mb{Q}_t)_b)- |z_b|^2(\mb{l}_t)_b\\
			%\label{eq:DF3}
			%\forall b = (i,j)\in\mcal{E}\\
			\label{eq:DF4}
			(\mb{v}_t)_i (\mb{l}_t)_b \geq (\mb{P}_t)_b^2 + (\mb{Q}_t)_b^2,
		\end{gather}
	\end{subequations}
	where $z_b \ldef r_b + jx_b$ is the impedance of branch $b\in\mcal{E}$. Note that~\eqref{eq:DF1}-\eqref{eq:DF2} hold for all $j\in\mcal{N}$, while~\eqref{eq:DF3}-\eqref{eq:DF4} hold for all branches $b=(i,j)\in\mcal{E}$.
	% 	\eqref{eq:DF1}-\eqref{eq:DF4} all hold for all time steps $t\in\mcal{T}$.
	
	We assume that ESS operators have provided vector $\mb{d}$ to the MGO \emph{a priori}. For a given $\mb{d}$, the LR problem uses the \emph{DistFlow} equations alongside other constrains to maximize restored load by setting load setpoints and charge/discharge amount corresponding to the decisions in $\mb{d}$. Because the exact load demand of the users may not be known \emph{a priori}, we use forecasts $\hat{\mb{p}},\hat{\mb{q}}\in\real{N_L T}$ in place of the actual demand values. To this end, we define the load \emph{pickup} vector $\mb{r}_t$ as
	\begin{gather}
		\label{eq:pickup}
		(\mb{r}_t)_i \ldef \frac{ (\mb{p}_t)_i}{(\hat{\mb{p}}_t)_j} \stackrel{[a]}{=} \frac{ (\mb{q}_t)_i}{(\hat{\mb{q}}_t)_j},\\
		\notag
		\text{index $i\in\mcal{N}$ corresponds to load bus $j\in[N_L]$},
	\end{gather}
	wherein $[a]$ follows from the assumption of constant power factor for all loads. If oversatisfaction of load demand is allowed, $\mb{r} \ldef \stack{\mb{r}_t}{\mcal{T}}$ takes values in $[0,c]^{N_L T}$, where $c\geq1$ denotes the maximum value of the pickup. In this work, we let $c=1$.
	
	Let $\mb{p}^E$ and $\mb{q}^E$ denote the sub-vectors of $\mb{p}$ and $\mb{q}$ corresponding to the ESS buses, respectively. Let $\mb{p}^\text{ch}$ and $\mb{p}^\text{dis}$ denote the ESS charge and discharge powers, respectively. Both vectors have the same size as $\mb{p}^E$. In addition, let $\mb{s}^\text{init}\in\real{N_E}$ represent the initial SoC vector. To this end, the load restoration problem is formulated as
	\begin{align}
		\label{eq:OPF}
		\mcal{LR}(\mb{d}) :=\mb{r}^*\in&  \argmax\limits_{\{\mb{r},\mb{p}^\text{ch},\mb{p}^\text{dis},\mb{p},\mb{q},\mb{P},\mb{Q},\mb{v},\mb{l},\mb{s}\}}  \pmb{\xi}^\top \mb{r}
	\end{align}
	\addtocounter{equation}{-1}
	\begin{subequations}
		\begin{align}
			\label{eq:OPF0}
			\text{s.t.} \quad& \text{\eqref{eq:DF1}-\eqref{eq:DF4}}\\
			\label{eq:OPF1}
			&\underline{\mb{v}} \leq \mb{v} \leq \bar{\mb{v}}\\
			\label{eq:OPF2}
			&\mb{0}_{ET} \leq \mb{l}  \leq\bar{\mb{l}}\\
			\label{eq:OPF3}
			&\mb{0}_{N_L T} \leq \mb{r} \leq \mb{1}_{N_LT}
		\end{align}
		\begin{gather}
			\label{eq:OPF4}
			(\mb{r}_t)_j \leq (\mb{r}_{t+1})_j, \forall j\in [N_L], \forall t \in [T-1]\\
			\label{eq:OPF5}
			0 \leq (\mb{p}^\text{ch}_t)_j \leq (\mb{d}_t)_j\bar{p}^\text{ch}, \forall j \in [N_E], \forall t \in \mcal{T}\\
			\label{eq:OPF6}
			0 \leq (\mb{p}^\text{dis}_t)_j \leq (1-(\mb{d}_t)_j)\bar{p}^\text{dis}, \forall j \in [N_E], \forall t \in \mcal{T}\\
			\label{eq:OPF7}
			\mb{p}^E = -\mb{p}^\text{ch} + \mb{p}^\text{dis}\\
			\label{eq:OPF8}
			\underline{\mb{q}}^E \leq \mb{q}^E \leq \bar{\mb{q}}^E\\
			\notag
			(\mb{s}_{t+1})_j = (\mb{s}_t)_j + \Gamma^\text{ch}(\mb{p}^\text{ch}_t)_j - \Gamma^\text{dis}(\mb{p}^\text{dis}_t)_j; \;\;\mb{s}_1 = \mb{s}^\text{init}\\
			\label{eq:OPF9}
			\forall j \in [N_E], \forall t \in \mcal{T}\\
			\label{eq:OPF10}
			\underline{\mb{s}} \leq \mb{s} \leq \bar{\mb{s}}.
		\end{gather}
	\end{subequations}
	The objective function in~\eqref{eq:OPF} maximizes the total pickup for all load buses, with each individual pickup weighted by the vector $\pmb{\xi}\in\real{N_LT}_{+}$. Constrains~\eqref{eq:OPF1}-\eqref{eq:OPF2} limit nodal voltages and line currents within their safe operational limits. Constrain~\eqref{eq:OPF3} restricts the pickups to $[0,1]$. Constrain~\eqref{eq:OPF4} ensures \emph{monotonic load restoration}, i.e. loads once picked up may not be dropped. Constrains~\eqref{eq:OPF5} and~\eqref{eq:OPF6} set limits for the ESS charge and discharge powers while ensuring \emph{complementarity}, i.e. the ESS may either charge or discharge on a given time step, but not both. Constrain~\eqref{eq:OPF7} prescribes that the total power output of ESSs equal the sum of charge and discharge powers, while constrain~\eqref{eq:OPF8} places bounds on the reactive power the ESS inverters can produce. Constrain~\eqref{eq:OPF9} describes the temporal evolution of ESS SoC, with $\Gamma^\text{ch}$ and $\Gamma^\text{dis}$ denoting charge power-to-SoC and SoC-to-discharge power conversion factors, respectively. Lastly, constrain~\eqref{eq:OPF10} restricts the SoC of all ESSs within their upper and lower bounds.
	
	Problem \eqref{eq:OPF} is a convexified AC-OPF problem that can be solved via off-the-shelf convex solvers. $\mcal{LR}$ is implemented at the MGC, which receives $\mb{d}$ from all ESSs and generates the optimal pickup $\mb{r}^*$ for all the loads. We assume that $\mb{d}$ is feasible for $\mcal{LR}(\mb{d})$, and we refer to such $\mb{d}$ as \emph{$\lr$-feasible}. In practice, the MGC can test feasibility of any provided $\mb{d}$, and reject infeasible ones. It is also assumed that the initial SoC vector $\mb{s}^\text{init}$ allows for at least one $\lr$-feasible value of $\mb{d}$.
	
	%
	% algorithms
	%
	\begin{figure}[!tb]
		\removelatexerror
		\SetKwRepeat{Do}{do}{while}
		\begin{algorithm}[H]
			\nonl
			\KwSty{Input:} mechanism $\lr$, $\mb{d}^\text{init}\in\mcal{D}^n$, tolerance $\emph{\texttt{tol}}>0$\\
			\nonl
			\KwSty{Output:} approx. $l_1$-sensitivity $\Delta_\lr$ of mechanism $\lr$\\
			\KwSty{set} $\mb{d}^{(0)} = \mb{d}^\text{init}$, $k = 0$, and $\Delta_\lr = -\infty$ \\
			\Do{$\left\{\left\lVert \lr(\mb{d}^{(k-1)}) - \lr(\mb{d}^{(k)}) \right\rVert_1\geq\texttt{tol}\right\}$}{
				\KwSty{set} $\delta_j^{(k)} \ldef \left\lVert \frac{\partial \lr(\mb{d}^{(k)})}{\partial(\mb{d}^{(k)})_j} \right\rVert_2, \;\; \forall j\in[N_ET]$\\
				%				\For{$j\gets 1$ \KwTo $N_ET$}
				%				{$\delta_j^{(k)} \ldef \left\lVert \frac{\partial \lr(\mb{d}^{(k)})}{\partial(\mb{d}^{(k)})_j} \right\rVert_2$}
				\KwSty{set} $j^* = \argmax\limits_{j\in[N_ET]} \left\{ \delta_j^{(k)} \right\}$\\
				{\KwSty{set} $(\tilde{\mb{d}}^{(k)})_i \ldef \begin{cases}
						(\mb{d}^k)_i & i\neq j^*\\
						1-(\mb{d}^k)_i & i = j^*
					\end{cases}, \;\; \forall i \in [N_ET]$}\\
				\eIf{$\tilde{\mb{d}}^{(k)}$ is $\lr$-feasible}{
					\KwSty{set} $\mb{d}^{(k+1)} = \tilde{\mb{d}}^{(k)}$\\
					\KwSty{define} $\phi\ldef \left\lVert \lr(\mb{d}^{(k+1)}) - \lr(\mb{d}^{(k)}) \right\rVert_1$\\
					\KwSty{if} $\Delta_\lr < \phi$ \KwSty{then set} $\Delta_\lr = \phi$\\}{
					\KwSty{set} $\mb{d}^{(k+1)} = \mathring{\mb{d}}$ where $\mathring{\mb{d}}$ satisfies $\widehat{\lr}(\tilde{\mb{d}}^{(k)}) = \lr(\mathring{\mb{d}})$\\}
				%				\If{ $\Delta_\lr < \delta^{(k)}_{j^*}$}
				%				{$\Delta_\lr = \delta^{(k)}_{j^*}$\\}
				\KwSty{set} $k = k+1$}
			\KwSty{return} $\Delta_\lr$
			\caption{Approximation of $\Delta_\lr$}
			\label{alg:sens_estimate}
		\end{algorithm}
	\end{figure}
	\section{Differentially Private Mechanism Design for $\mcal{LR}(\mb{d})$}
	In this section, we discuss various aspects of making the LR problem deferentially private.
	\subsection{Trust Architecture}
	To implement DP in a real system, identification of trusted as well as potentially adversarial agents within the system becomes paramount. In the context of the LR problem, we assume that the MGC, operated by the MGO, is a trusted arbiter for all MG participants. Furthermore, we assume that all problem parameters of~\eqref{eq:OPF} (except $\mb{d}$) are known to all participants in the MG. However, vector $\mb{d}$ is supposed to be the private data of the ESS operators. To protect the privacy of $\mb{d}$ by avoiding its inference from the optimal pickup $\mb{r}^*$, the MGC adds noise to it as stipulated by DP. This can be represented by the mechanism $\mcal{M}^\lr(\mb{d}) \ldef \lr(\mb{d}) + \mb{w}$. From Theorem~\ref{th:lap-dp}, we see that for $\varepsilon>0$, letting $(\mb{w})_i \sim \lap{\frac{\Delta_\lr}{\varepsilon}}$ makes the mechanism $\mcal{M}^\lr$ $\varepsilon$-DP, where 
	\begin{align}
		\label{eq:L1sens}
		\Delta_\lr \ldef \max\limits_{\breve{\mb{d}},\breve{\mb{d}}'\in\{0,1\}^{N_ET}:(\breve{\mb{d}}\sim_1\breve{\mb{d}}')} \spnorm{ \lr(\breve{\mb{d}}) - \lr(\breve{\mb{d}}') }{1}.
	\end{align}
	The resulting $\tilde{\mb{r}} := \mcal{M}^\lr(\mb{d})$ may not be $\lr$-feasible. Thus, we run a post-processing procedure $\mcal{FR}(\tilde{\mb{r}})$ which yields a new pickup $\hat{\mb{r}}$, and a corresponding $\lr$-feasible $\hat{\mb{d}}$. The MGC then withdraws power from the ESSs according to $\hat{\mb{d}}$ and restores load according to $\hat{\mb{r}}$. By Theorems~\ref{th:pp} and~\ref{th:lap-dp}, $\mcal{FR}(\mcal{M}^\lr(\mb{d}))$ is $\varepsilon$-DP. Next,
	we will discuss the calculation of $\Delta_\lr$ and the feasibility restoration operator $\mcal{FR}$.
	
	\begin{figure}[!tb]
		\removelatexerror
		\begin{algorithm}[H]
			\label{alg:feas_restore}
			\caption{$\widehat{\lr}(\mb{d})$: Generation of $\lr$-Feasible Mode Vector}
			\nonl
			\KwSty{Input:} mode vector $\mb{d}$\\
			\nonl
			\KwSty{Output:} $\widehat{\lr}(\mb{d}) \ldef \lr(\mb{d}^{(u)})$ where $\mb{d}^{(u)}$ is $\lr$-feasible\\
			\KwSty{set} $\mb{d}^{(0)} = \mb{d}$ and $u = 0$\\
			\While{$\mb{d}^{(u)}$ is not $\lr$-feasible}
			{randomly select $j\in[N_ET]$ under uniform distribution
				% \tcc*[f]{each $j\in[N_ET]$ should have equal probability of getting selected}\\
				$(\mb{d}^{(u+1)})_i \ldef \begin{cases}
					(\mb{d}^{(u)})_i & i \neq j\\
					1-(\mb{d}^{(u)})_i & i = j
				\end{cases}, \;\; \forall i \in [N_ET]$\\
				\KwSty{set} $u = u+1$
			}
			\KwSty{return} $\lr(\mb{d}^{(k)})$
		\end{algorithm}
	\end{figure}
	\subsection{Calculation of $\Delta_{\lr}$}
	Due to the structure of the pickup vector, we have  $\Delta_\lr \leq N_L T$. However, setting $\Delta_\lr=N_LT$ is excessively conservative since this would require the existence of $\mb{d}' \sim_1 \mb{d}''$ such that $\lr(\mb{d}')=\mb{0}_{N_LT}$ and $\lr(\mb{d}'')=\mb{1}_{N_LT}$, which may not exist. Therefore, we propose a heuristic for calculation of $\Delta_\lr$ in Algorithm~\ref{alg:sens_estimate}. Given the mechanism $\lr$ and $\mb{d}^\text{init}$, the proposed algorithm generates a sequence of $\{ \mb{d}^{(k)} \}_k$ wherein for a given $\mb{d}^{(k)}$, $\mb{d}^{(k+1)}$ is generated such as to increase the value of $h(\mb{d}^{(k)},\mb{d}^{(k+1)})\ldef\spnorm{\lr(\mb{d}^{(k)}) - \lr(\mb{d}^{(k+1)})}{1}$ while trying to ensure that $\mb{d}^{(k)}\sim_1 \mb{d}^{(k+1)}$. This is done by calculating $\delta^{(k)}_j$, which is the sensitivity of $\lr(\mb{d}^{(k)})$ to the $\tth{j}$ index of $\mb{d}^{(k)}$, for all $j\in[N_ET]$. This is followed by generation of an intermediate vector $\tilde{\mb{d}}^{(k)}$ by flipping the $\tth{j^*}$ bit of $\mb{d}^{(k)}$ where $j^*$ is the index that maximizes $\delta^{(k)}_j$. If $\tilde{\mb{d}}^{(k)}$ is $\lr$-feasible, then $\mb{d}^{(k+1)}$ is set equal to $\tilde{\mb{d}}^{(k)}$ and $h(\mb{d}^{(k)},\mb{d}^{(k+1)})$ is tested as a candidate for $\Delta_\lr$. If $\tilde{\mb{d}}^{(k)}$ is not $\lr$-feasible, Algorithm ~\ref{alg:feas_restore} is used to convert $\tilde{\mb{d}}^{(k)}$ into a $\lr$-feasible $\mb{d}^{(k+1)}$ by randomly flipping bits at various indices and checking for feasibility. However, $h(\mb{d}^{(k)},\mb{d}^{(k+1)})$ is not tested as a candidate for $\Delta_\lr$ since $\mb{d}^{(k)}\sim_1 \mb{d}^{(k+1)}$ does not hold.
	
	\subsection{Feasibility Restoration Operator $\mcal{FR}(\mb{r})$}
	For a given $\mb{r}$, the feasibility restoration operator $\mcal{FR}(\mb{r})$ is defined by the following problem. It calculates the minimum-norm perturbation $\hat{\pmb{\rho}}\in\real{N_LT}$ such that $\mb{r}+\hat{\pmb{\rho}}$ is feasible, i.e. $\lr(\hat{\mb{d}})=\mb{r}+\hat{\pmb{\rho}}$ for some mode vector $\hat{\mb{d}}$. Correspondingly, $\mb{r}+\hat{\pmb{\rho}}$ is the actual pickup delivered by the MGC.
	\begin{align}
		\label{eq:FR}
		\mcal{FR}(\mb{r}) \ldef  \hat{\mb{d}},\hat{\pmb{\rho}}\in \argmin\limits_{\mb{d},\pmb{\rho},\mb{p}^\text{ch},\mb{p}^\text{dis},\mb{p},\mb{q},\mb{P},\mb{Q},\mb{v},\mb{l},\mb{s}} \spnorm{\pmb{\rho}}{2}
	\end{align}
	\addtocounter{equation}{-1}
	\begin{subequations}
		\begin{align}
			\label{eq:FR1}
			\text{s.t. }& \text{\eqref{eq:OPF0}-\eqref{eq:OPF2},~\eqref{eq:OPF5}-\eqref{eq:OPF10}}\\
			\label{eq:FR2}
			&\mb{0}_{N_L T} \leq \mb{r} + \pmb{\rho} \leq \mb{1}_{N_LT}\\
			\label{eq:FR3}
			&(\mb{r}_t)_j + (\pmb{\rho}_t)_j \leq (\mb{r}_{t+1})_j+(\pmb{\rho}_{t+1})_j,\\
			\notag
			& \forall j\in [N_L], \forall t \in [T-1]\\
			\label{eq:FR4}
			& \mb{d} \in \{0,1\}^{N_ET}.
		\end{align}
	\end{subequations}
	\eqref{eq:FR} is a mixed-integer program, which may become intractable as the number of binary variables increases. However, many practical workarounds have been devised for efficiently solving mixed-integer problems, including branch-and-bound methods and continuous reformulations~\cite{SAG-ML-MS:2021}, which can be used for efficiently computing $\mcal{FR}(\mb{r})$.
	
	Note that for $\hat{\mb{d}},\hat{\pmb{\rho}}=\mcal{FR}(\mcal{M}^\lr(\mb{d}))$, in general $\hat{\mb{d}}\neq \mb{d}$, i.e. the mode vector implemented by the MGC is different from the one requested by ESS operators. This is a consequence of post-processing with $\mcal{FR}$ which cannot use $\mb{d}$ to protect its privacy, and must instead only use $\mcal{M}^\lr(\mb{d})$. Calculation of probabilistic guarantees on the accuracy of $\hat{\mb{d}}$ with respect to $\mb{d}$ is out of scope of the present work.
	
	\section{Simulation Results}
	In this section, we demonstrate the performance of differentially private LR through simulations. We use the $33$-bus radial feeder test case `\texttt{case33bw}' in \texttt{MATPOWER}~\cite{RDZ-etal:2011}, which prescribes network topology, branch parameters, and load demand, to emulate the MG. We consider a scenario with 7 ESSs having identical system parameters, which (alongside other values) are listed in Table~\ref{tab:params}.
	\begin{table}[!tb]
		\centering
		\caption{Simulation Setup: System parameters}
		\label{tab:params}
		\begin{tabular}{|c|c|}
			\hline
			\textbf{Parameter} & \textbf{Value}\\
			\hline
			$N_E$, $N_L$, $N$, $E$ & $7$, $26$, $33$, $32$\\
			\hline
			$\underline{v}$, $\bar{v}$, $\bar{l}$ & $0.9$ p.u, $1.1$ p.u, $\infty$ (for all buses and lines)\\
			\hline
			$\mcal{N}^\text{ESS}$ & $\{ 2,7,12,17,23,27,31 \}$\\
			\hline
			$\underline{s}$, $\bar{s}$, $\bar{p}^\text{ch}$, $\bar{p}^\text{dis}$ & $0.3594$MWh, $3.5940$MWh, $1.1980$MW, $1.1980$MW\\
			\hline
			$\Gamma^\text{ch}$, $\Gamma^\text{dis}$ & $0.90$h, $1.11$h\\
			\hline
		\end{tabular}
	\end{table}
	For convex problems and the mixed-integer counterparts, the \texttt{Gurobi} solver~\cite{gurobi} was used to find a solution, interfaced with \texttt{MATLAB} through \texttt{CVX}~\cite{cvx}. For the system under consideration, Algorithm~\ref{alg:sens_estimate} yielded the value $\Delta_{\lr} = 1.2863$. The generated mode vectors $\mb{d}^{(k)}$  were unique up to $k=12$, and then started oscillating between two values on successive time steps. On no iteration $k$ did the vector $\mb{d}^{(k)}$ fail to be $\lr$-feasible, and therefore the mechanism $\widehat{\lr}$ was never called.
	
\begin{figure}[!tb]
		\centering
		\setkeys{Gin}{width=0.5\linewidth}
		\subfloat[\label{fig:1a}]{\includegraphics{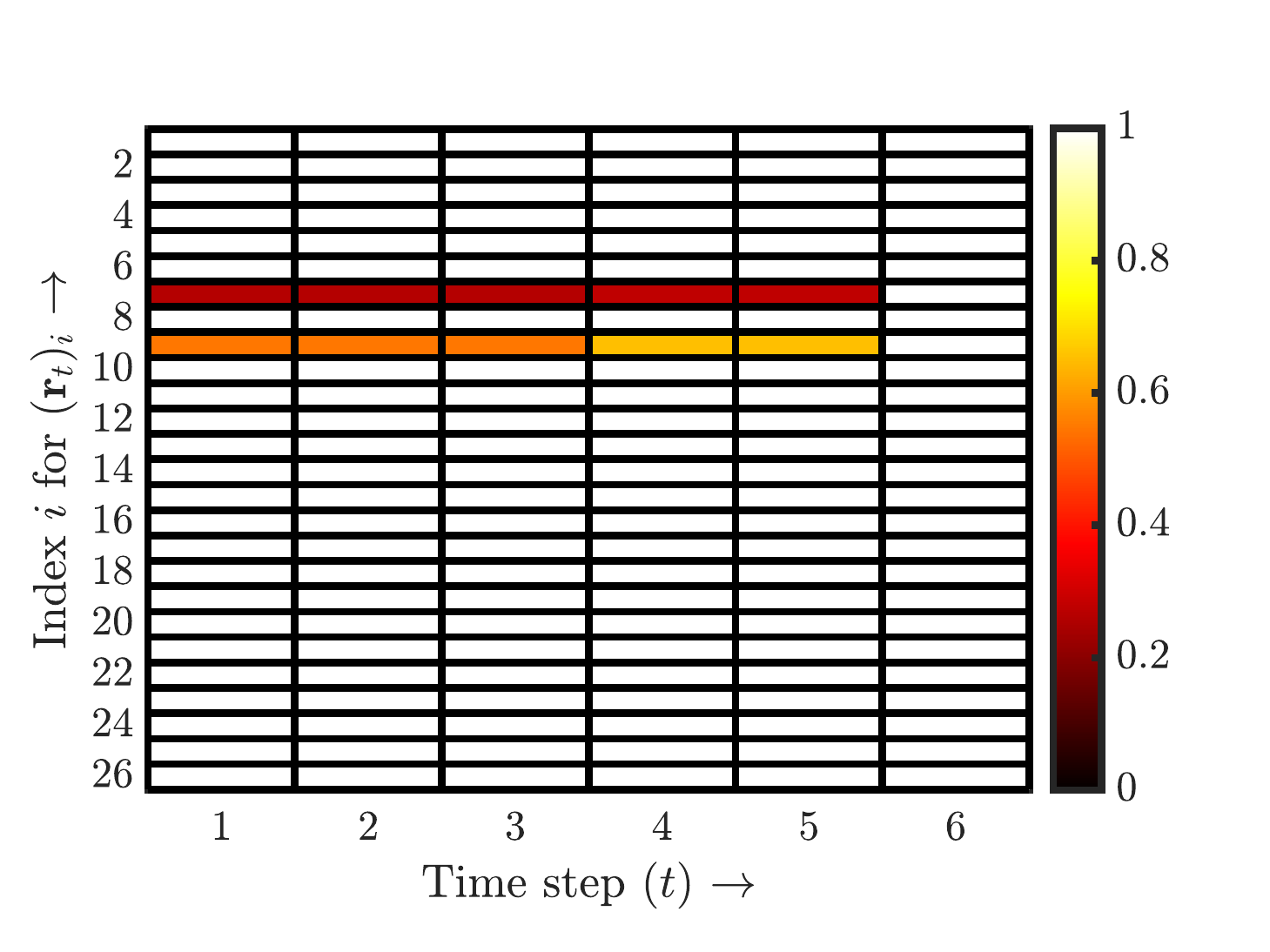}}
		\hfill
		\subfloat[\label{fig:1b}]{\includegraphics{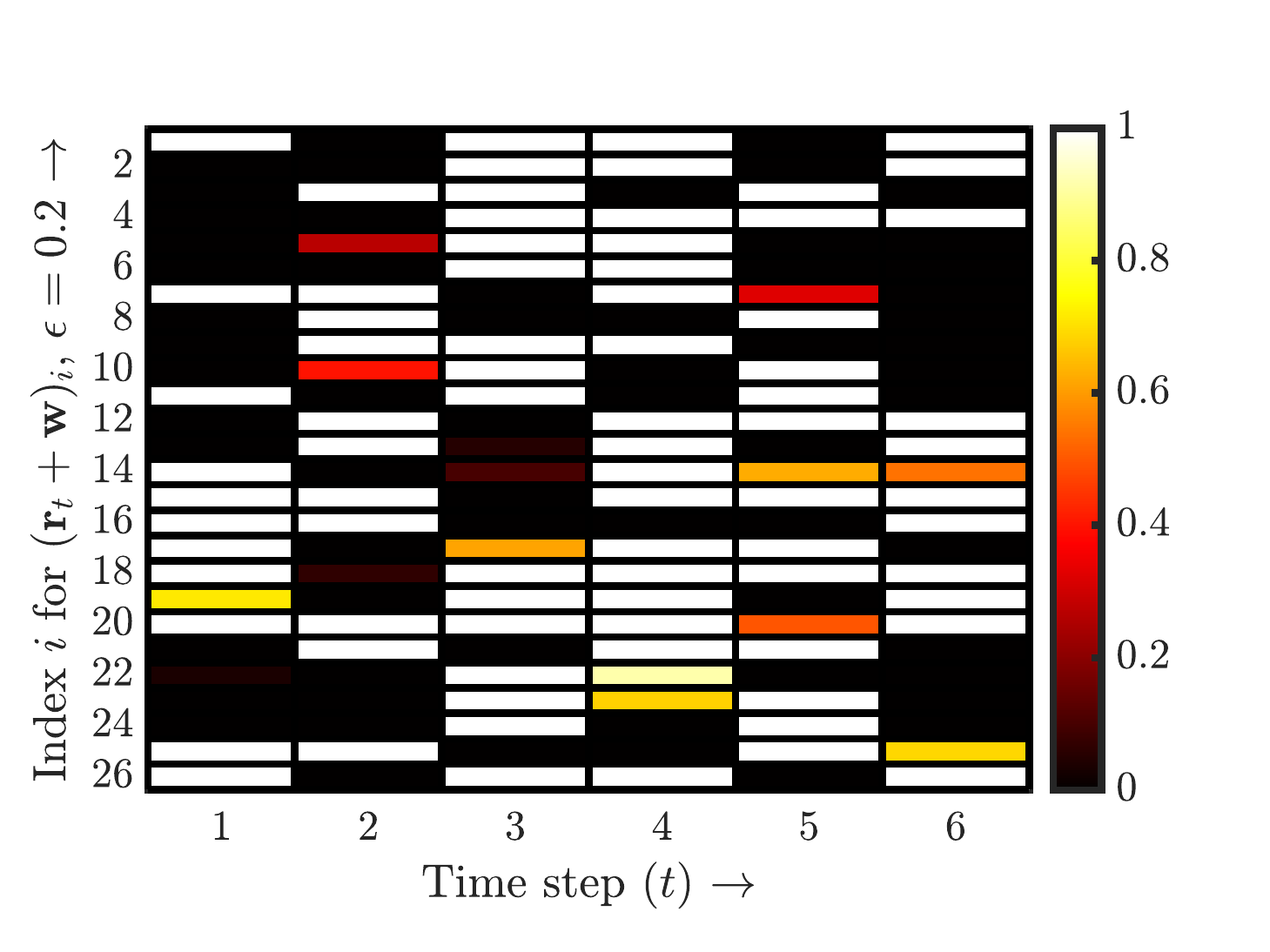}}
		\hfill
		\subfloat[\label{fig:1c}]{\includegraphics{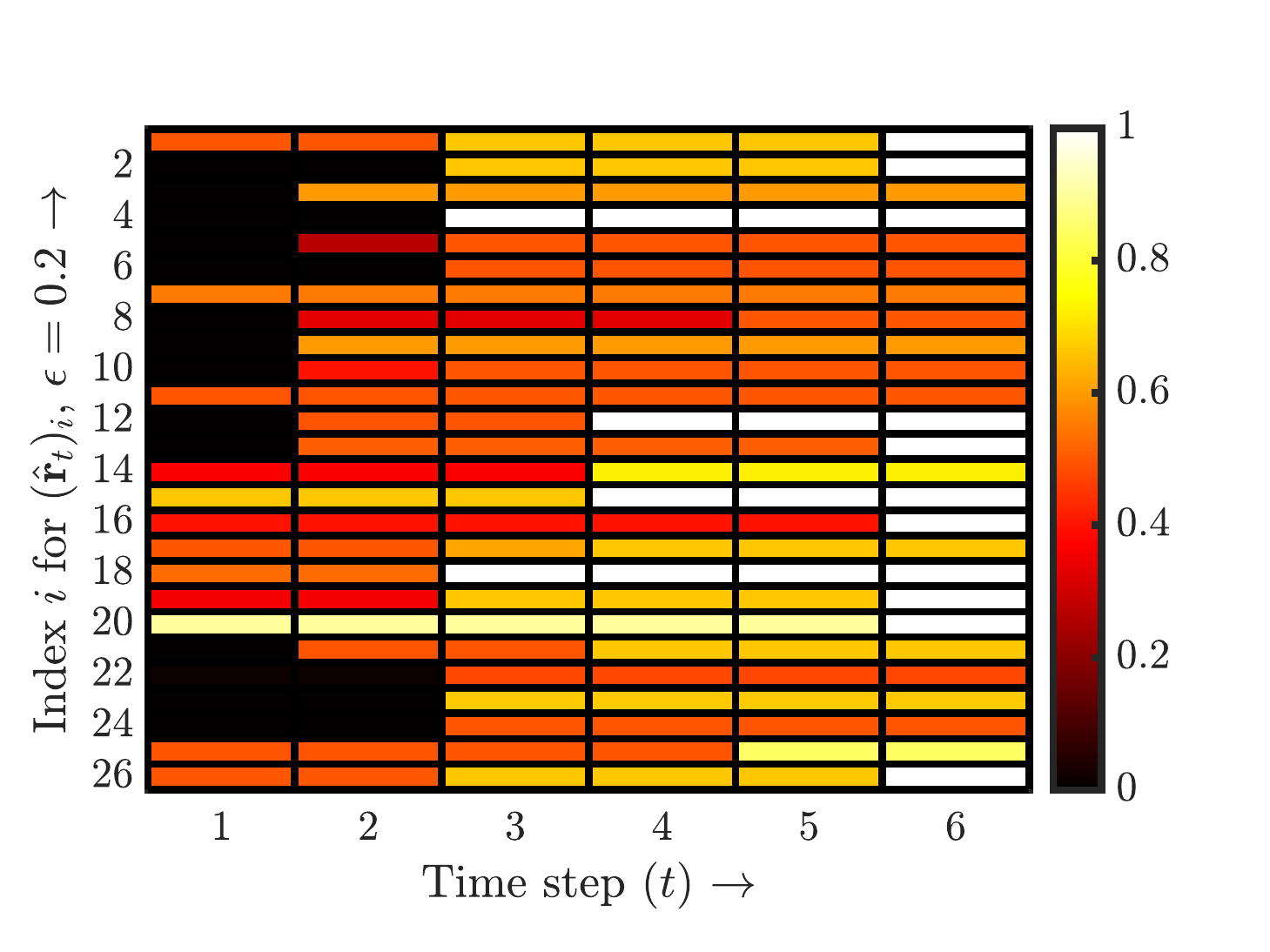}}
		\hfill
		\subfloat[\label{fig:1d}]{\includegraphics{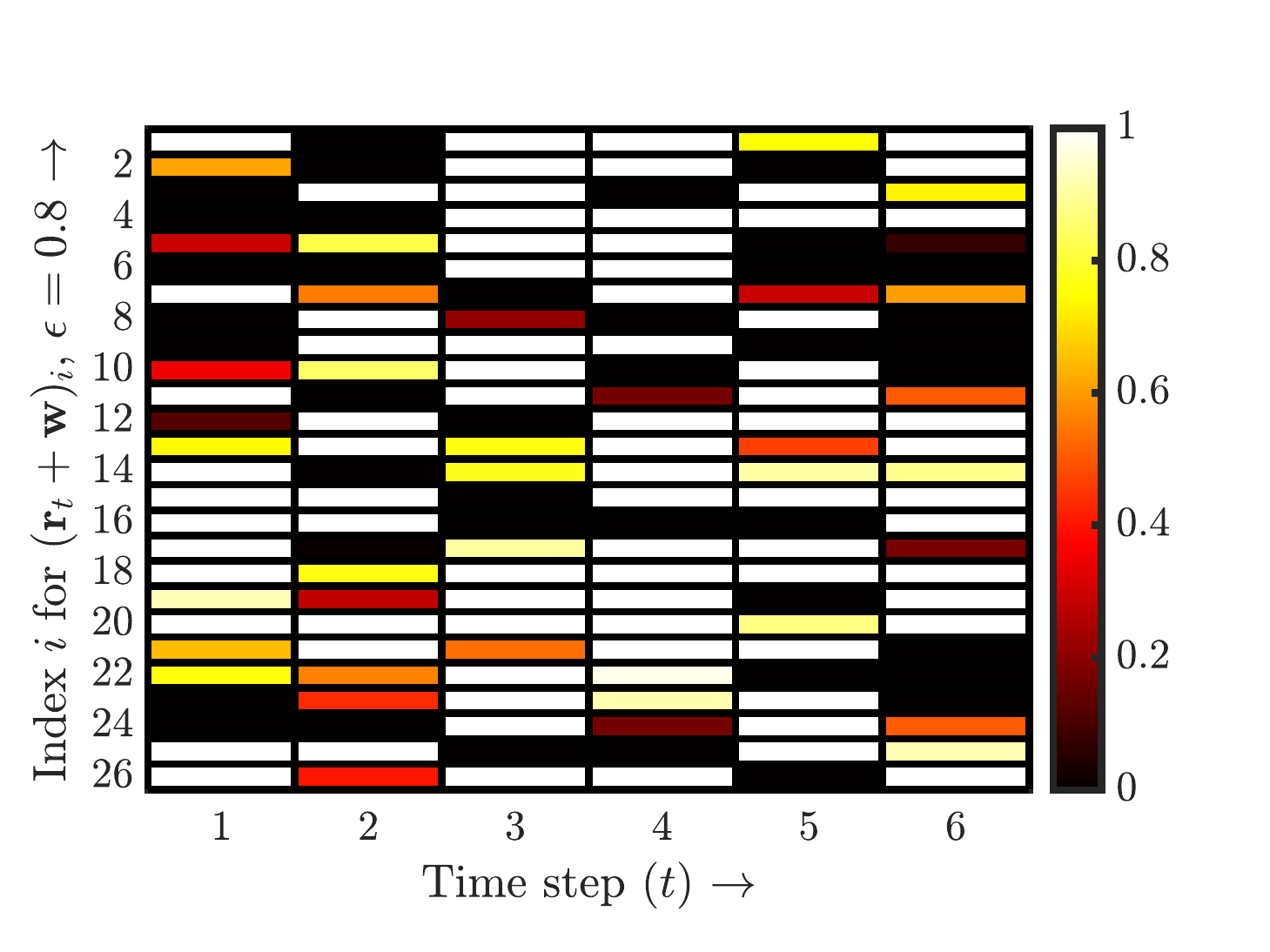}}
		\hfill
		\subfloat[\label{fig:1e}]{\includegraphics{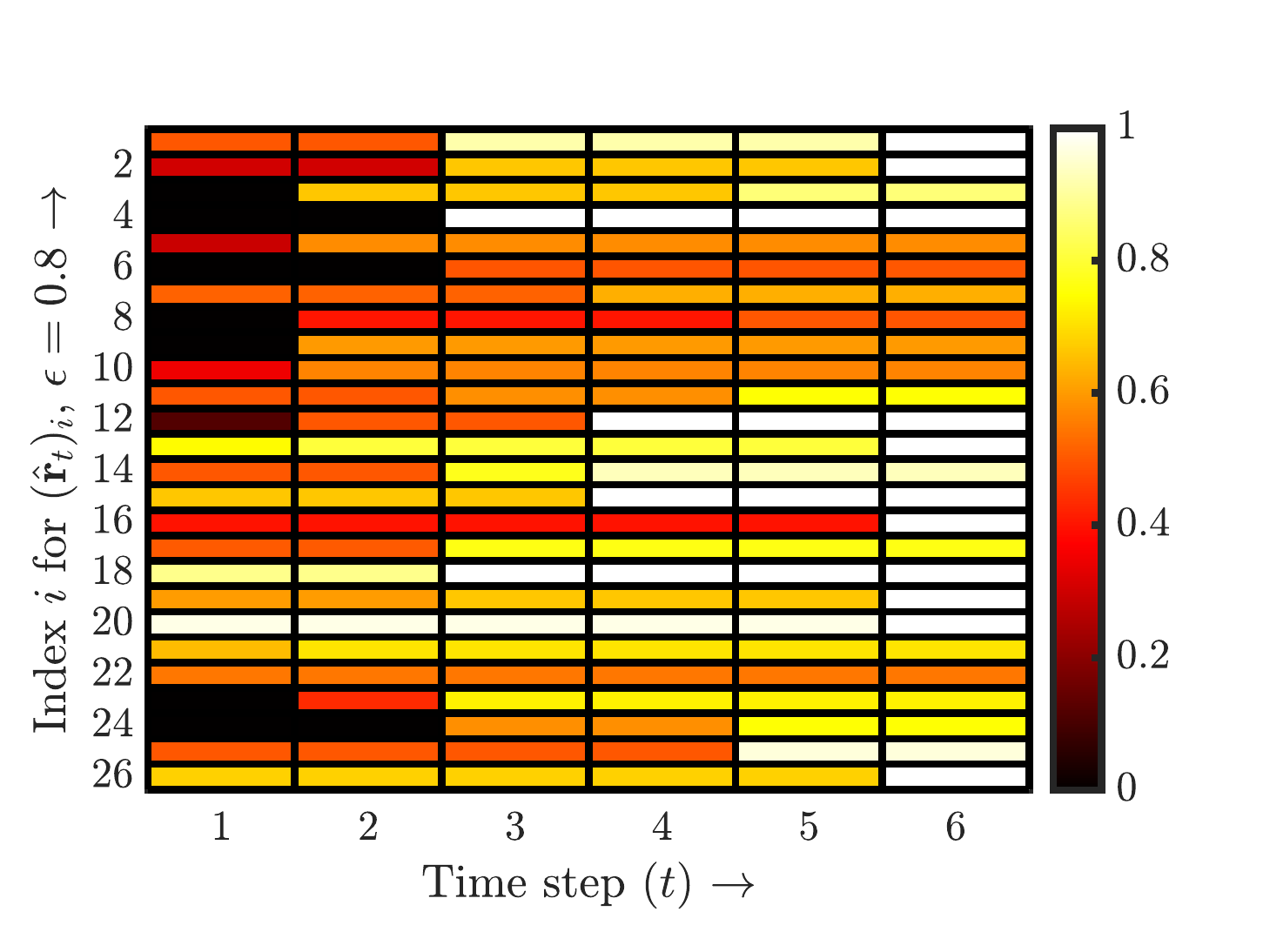}}
		\hfill
		\subfloat[\label{fig:1f}]{\includegraphics{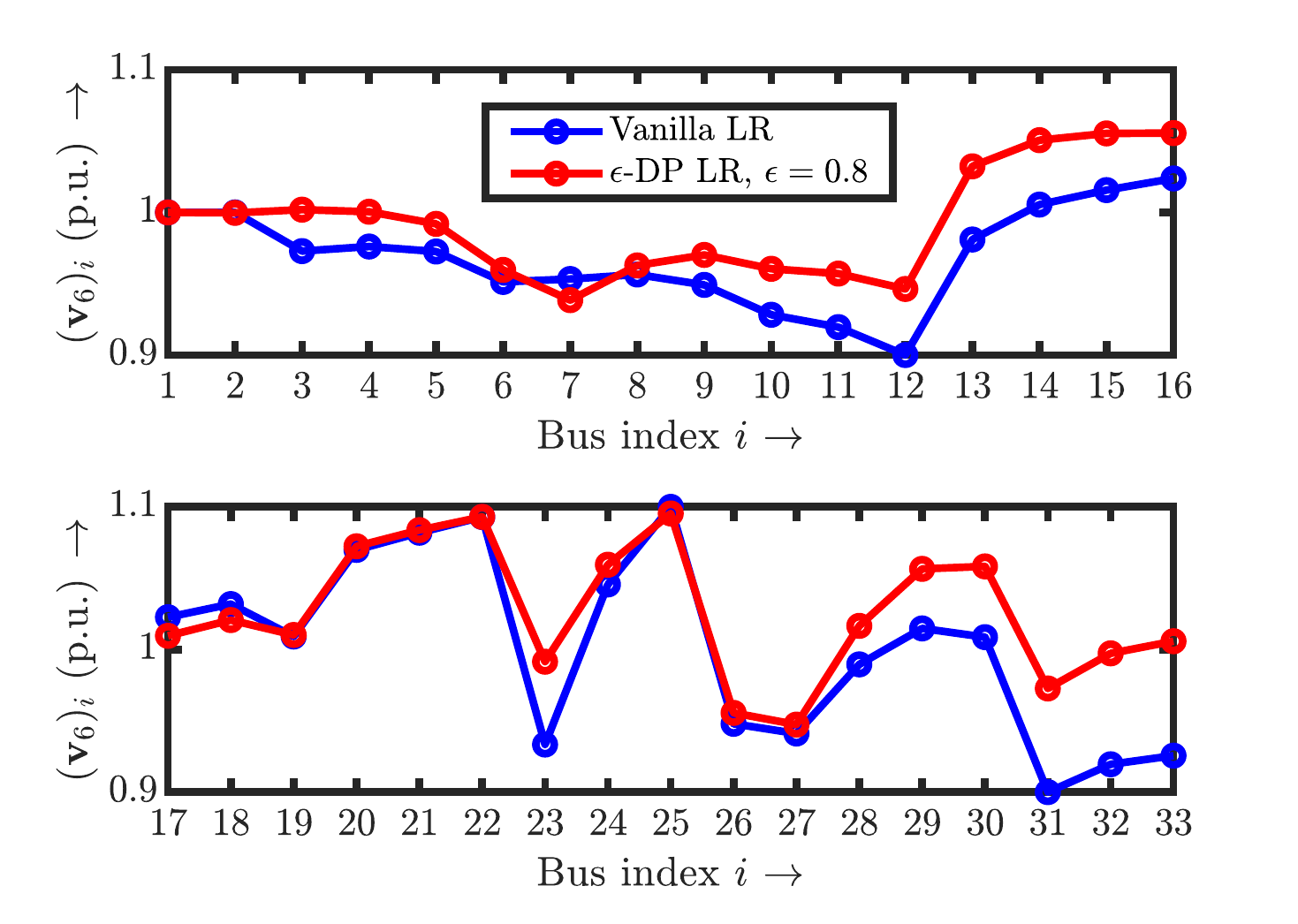}}
		\caption{The comparisons between non-private LR and DP-LR for different values of $\varepsilon$: (a) Pickup values for non-private LR; (b) and (d) Noisy pickup values before post-processing for $\varepsilon=0.2, 0.8$, respectively; (c) and (e) Post-processed pickup values for $\varepsilon=0.2, 0.8$, respectively; and (f) Nodal voltages for non-private LR and DP-LR for $\varepsilon=0.8$.}
	\end{figure}

	We carry out simulation over 6 time steps, i.e. $T=6$. Every element of $\mb{s}^\text{init}$ is chosen uniformly at random from $[0.7\bar{s},0.9\bar{s}]$ and then fixed for all simulations. The initial $\mb{d}$ is chosen from a binomial distribution with parameter $0.4$. Figure~\ref{fig:1a} shows the pickup values for non-private LR, with most loads on all time steps receiving full load satisfaction. Figures~\ref{fig:1b} and~\ref{fig:1d} show the noisy pickup values for $\varepsilon=0.2$ and $\varepsilon=0.8$, respectively, while Figures~\ref{fig:1c} and~\ref{fig:1e} show the post-processed pickup values for said values of $\varepsilon$. Comparing Figures~\ref{fig:1b} and~\ref{fig:1d}, it is evident that a lower value of $\varepsilon$ leads to addition of more noise to the LR output, thereby providing better privacy protection. The flip side of noise addition can be seen in the suboptimality of LR in the DP setting. Compared to Figure~\ref{fig:1a}, the pickup of loads in Figures~\ref{fig:1c} and~\ref{fig:1e} is lower, and this loss of LR performance can be viewed as a trade-off for privacy of the ESS mode vector. Furthermore, comparison of Figure~\ref{fig:1b} with Figure~\ref{fig:1c} (and Figure~\ref{fig:1d} with Fig~\ref{fig:1e}) clearly demonstrates the effectiveness of post-processing, as the post-processed pickups show monotonic load restoration. By contrast, the pickups without post-processing do not demonstrate the same. For $\varepsilon=0.2$, the initial $\mb{d}$ differed from the implemented $\hat{\mb{d}}$ in $29$ out of $42$ indices, while for $\varepsilon=0.8$ they differed in $24$ indices. This shows empirically that lower values of $\varepsilon$ lead to a higher deviation of $\hat{\mb{d}}$ from $\mb{d}$.
	
	Figure~\ref{fig:1f} compares the nodal voltage quantities under non-private LR with that of DP-LR on the last time step ($t=6$). Again, the effectiveness of post-processing can be clearly seen, as the voltage profile in both cases respect the voltage upper and lower bounds over all time steps. 
	
	\section{Conclusion}
	In this paper, we considered the emerging issue of differential privacy in load restoration for multi-user MGs. We designed a differentially private mechanism which ensures privacy of ESS charge/discharge mode by calculating the optimal pickup vector and then perturbing it with carefully chosen noise. We further proposed a post-processing step to restore feasibility, and validated our results through simulations. The simulations demonstrated empirically that there exists a tradeoff between privacy and LR performance. Topics of further research involve finding bounds on the accuracy of the proposed mechanism, and results which exploit the network structure and parameters.


\end{document}